\documentclass[a4paper,12pt]{article}
\usepackage{epsfig}

\newcommand{\be}{\begin{equation}}
\newcommand{\ee}{\end{equation}}
\newcommand{\ba}{\begin{eqnarray}}
\newcommand{\ea}{\end{eqnarray}}
\newcommand{\ban}{\begin{eqnarray*}}
\newcommand{\ean}{\end{eqnarray*}}

\newcommand{\demi}{\frac{1}{2}}

\begin{document}

\title{\Large\bf  Quantum mechanical 'Backward in Time'?\\Comments, Answers, and\\
the Causal Indistinguishability Condition}

\author{{\bf Antoine Suarez}\thanks{suarez@leman.ch}\\ Center for Quantum
Philosophy\\ The Institute for Interdisciplinary Studies\\ P.O. Box
304, CH-8044 Zurich, Switzerland}


\maketitle

\vspace{2cm}
\begin{abstract}
We discuss a number of comments on quant-ph/9801061, and propose to
introduce the concept of 'Causal Indistinguishability'. The
incompatibility between Quantum Mechanics and Nonlocal Causality
appears to be unavoidable: upholding of Quantum Mechanics by
experiment would mean to live with influences backward in time,
just as we are now living with such faster than light.\\


\end{abstract}

\pagebreak

\begin{figure}[t]
\centering\epsfig{figure=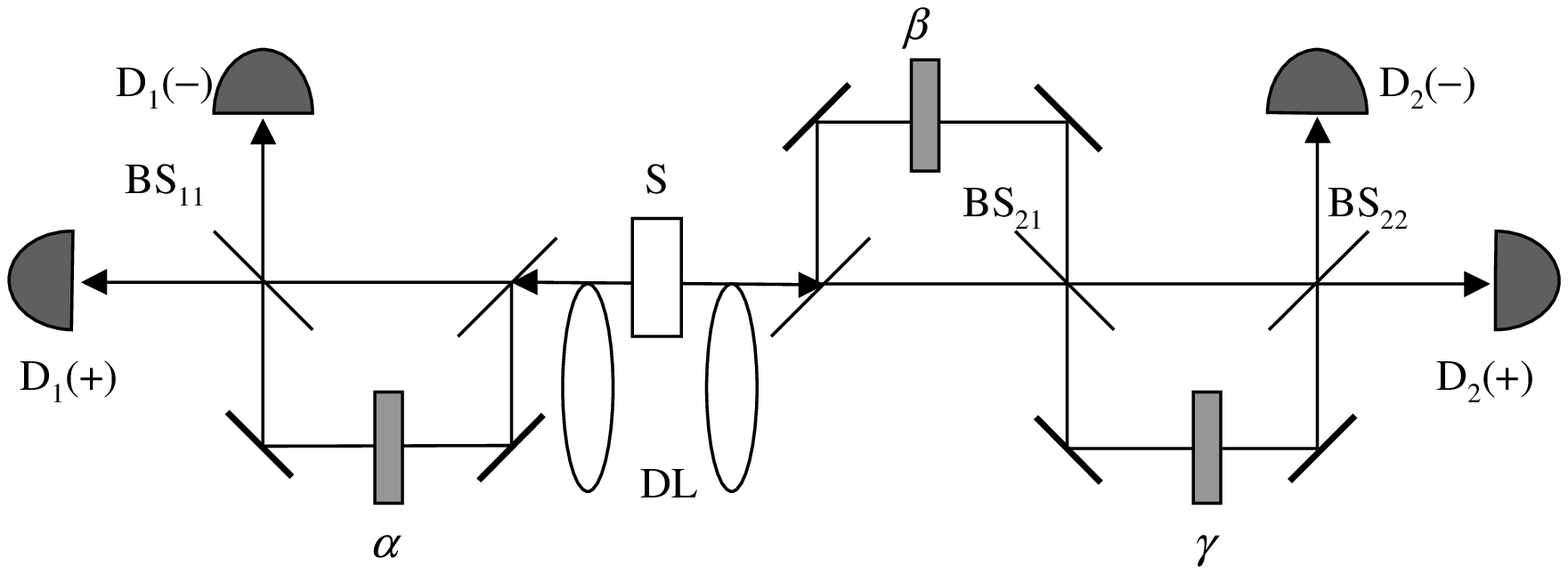,width=120mm}
{\caption{\footnotesize Impact series experiment with photon pairs.
See text for detailed description.}}
\label{fig:COMfig1}
\end{figure}

\bigskip

\section {Introduction}

In two recent papers \cite{as98.1, as98.2} impact series
experiments using the setup represented in Fig. 1 have been
proposed, and argued the quantum mechanical superposition principle
to imply influences acting backward in time, and therefore to be at
odds with the causality principle, and, in particular, with the
multisimultaneous causal view. For the sake of convenience we
present once again in section 2 the quantum mechanical view. In
section 3 we discuss the main comments received to date, most of
them as private communications and at the occasion of seminar
presentations. This discussion brings forward the concept of {\em
causal indistinguishability}, which is used in section 4 to improve
the multisimultaneous causal description \cite{asvs97.1, as97.2,
asvs97.2, as97.1}. A possible real experiment is discussed in
section 5, and in section 6 it is concluded that the conflict
between Quantum Mechanics and Causality cannot be escaped.\\

\section{The quantum mechanical view}

Consider again the setup sketched in Fig.1. Photon 1 enters the
left hand side interferometer and impacts on beam-splitter
BS$_{11}$ before being detected in either D$_{1}(+)$ or D$_{1}(-)$,
while photon 2 enters the 2-interferometer series on the right hand
side impacting successively on BS$_{21}$ and BS$_{22}$ before being
detected in either D$_{2}(+)$ or D$_{2}(-)$. Each interferometer
consists in a long arm of length $L$, and a short one of length
$l$. We assume the path difference $L-l$ set to a value which
largely exceeds the coherence length of the photon pair light, but
which is still smaller than the coherence length of the pump laser
light. For a pair of photons, eight possible path pairs lead to
detection. We label them as follows: $(l,ll)$; $(L,ll)$; $(l,Ll)$
and so on; where, e.g., $(l,Ll)$ indicates the path pair in which
photon 1 has taken the short arm, and photon 2 has taken first the
long arm, then the short one. By means of delay lines DL different
time orderings in the laboratory frame can be arranged.\\

We distribute the ensemble of all possible paths in the four
following subensembles:\\

\be
\begin{array}{lll}
(l,LL)&:&2L-l\\
(L,LL)\,,\,(l,Ll)\,,\,(l,lL)\,&:&L\\
(l,ll)\,,\,(L,Ll)\,,\,(L,lL)\,&:&l\\
(L,ll)&:&2l-L
\end{array}
\label{eq:paths}
\ee

where the right-hand side of the table indicates the path
difference between the single paths of each photon characterizing
each subensemble of path pairs.\\

By means of delay lines DL different time orderings can be
arranged. We are mainly interested in the following two:

\begin{enumerate}
\item{{\em Time ordering 1:} The impact on BS$_{22}$ and detection at D$_{2}(\omega)$,
lie time-like separated before the impact on BS$_{11}$.}
\item{{\em Time ordering 2:} The impact on BS$_{11}$ and detection at D$_{1}(\sigma)$,
lie time-like separated before the impact on BS$_{21}$.}
\end{enumerate}

According to the conventional quantum mechanical definition of
indistinguishability the three paths belonging to subensemble $L$
have to be considered indistinguishable, and the same holds for the
three paths belonging to subensemble $l$. Moreover ordinary QM
assumes indistinguishability to be a sufficient condition for
observing quantum interferences and entanglement. Therefore the
superposition principle states for any possible time ordering:

\ba
P^{QM}_{\sigma\omega}(L)=|A_{\sigma\omega}(L,LL)+A_{\sigma\omega}(l,Ll)+
A_{\sigma\omega}(l,lL)|^2
\label{eq:JPQML}
\ea

where $P^{QM}_{\sigma\omega}(L)$ \,($\sigma,\omega\in\{+,-\}$)
denotes the joint probability of getting the outcome specified in
the subscript for a path of subensemble $L$, and
$A_{\sigma\omega}(path)$ the corresponding probability amplitudes
for the path specified in the parenthesis.\\

Substituting the amplitudes given in (\ref{Al,Ll}), (\ref{Al,lL})
and (\ref{AL,LL}) of the Appendix into Eq. (\ref{eq:JPQML}) yields
the following values for the conventional joint probabilities:

\ba
P^{QM}_{+ +}(L)&=&\frac{1}{12}
\Big[3-2\cos(\alpha+\beta)-2\cos(\alpha+\gamma)+2\cos(\gamma-\beta)\Big]\nonumber\\
P^{QM}_{+ -}(L)&=&\frac{1}{12}
\Big[3-2\cos(\alpha+\beta)+2\cos(\alpha+\gamma)-2\cos(\gamma-\beta)\Big]\nonumber\\
P^{QM}_{- +}(L)&=&\frac{1}{12}
\Big[3+2\cos(\alpha+\beta)+2\cos(\alpha+\gamma)+2\cos(\gamma-\beta)\Big]\nonumber\\
P^{QM}_{- -}(L)&=&\frac{1}{12}
\Big[3+2\cos(\alpha+\beta)-2\cos(\alpha+\gamma)-2\cos(\gamma-\beta)\Big]
\label{eq:qmjp}
\ea

From (\ref{eq:qmjp}) one is led to the following correlation
coefficient:

\ba
E^{QM}=\sum_{\sigma,\omega}(-\sigma\omega)
P^{QM}_{\sigma\omega}(L)&=&\frac{2}{3}\cos(\alpha+\gamma),
\label{eq:qmcc}
\ea

\smallskip

We are also interested in the single probabilities at each side of
the setup, i.e., the probability of getting a count in detector
D$_{2}(\sigma)$ independently of where photon 1 is detected, which
we denote $P^{QM}_{\pm\sigma}(L)$, and the probability of getting a
count in detector D$_{1}(\sigma)$ independently of where photon 2
is detected, which we denote $P^{QM}_{\sigma\pm}(L)$.\\

The single probabilities are related to the conventional joint ones
as follows:

\be
P^{QM}_{\pm\omega}(L)\,\equiv\,P^{QM}_{+\,\omega}(L)+P^{QM}_{-\,\omega}(L),
\label{eq:SPJP1}
\ee

and

\be
P^{QM}_{\sigma\pm}(L)\,\equiv\,P^{QM}_{\sigma\,+}(L)+P^{QM}_{\sigma\,-}(L).
\label{eq:SPJP2}
\ee

\smallskip

Eq. (\ref{eq:qmjp}) leads to the corresponding single probabilities
for the detections at side 2 (right-hand side) of the setup:

\ba
P^{QM}_{\pm+}(L)=\demi+\frac{1}{3}\cos(\beta-\gamma)\nonumber\\
P^{QM}_{\pm-}(L)=\demi-\frac{1}{3}\cos(\beta-\gamma),
\label{eq:SPQML1}
\ea

and at side 1 (left-hand side) of the setup:

\ba
P^{QM}_{+\pm}(L)=\demi-\frac{1}{3}\cos(\alpha+\beta)\nonumber\\
P^{QM}_{-\pm}(L)=\demi+\frac{1}{3}\cos(\alpha+\beta)
\label{eq:SPQML2}
\ea

\smallskip

Since in the impact series experiment we are considering detections
at one side of the setup occur time-like separated after the
detections at the other one, which measurement is made first and
which after does not depend at all on any inertial frame.
Therefore, in agreement with the principle that the effects cannot
exist before the causes, it is reasonable to assume that the
correlations appear because the photon impacting first chooses its
outcome without being influenced by the parameters the photon
impacting later will meet at the other arm of the setup, and the
photon impacting later chooses its outcome taking account of the
choice the photon impacting first has made.\\

The quantum mechanical predictions (Eq.
(\ref{eq:SPQML1}),(\ref{eq:SPQML2})) respect this causality view
for Time ordering 1, but clearly violate it for Time ordering 2,
and therefore one has to conclude that QM implies influences acting
backward in time between time-like separated regions.\\

Could such a retrocausation effect be used to built a time machine?
Consider the single probabilities for the subensemble with path
difference $l$ in Table (\ref{eq:paths}). The superposition
principle of QM states:

\ba
P^{QM}_{\sigma\omega}(l)&=&
\left|A_{\sigma\omega}(l,ll)+A_{\sigma\omega}(L,Ll)+
A_{\sigma\omega}(L,lL)\right|^2
\label{eq:proba3}
\ea

Substituting the amplitudes of (\ref{Al,ll}), (\ref{AL,lL}) and
(\ref{AL,Ll}) in the Appendix into Eq. (\ref{eq:proba3}) one gets:

\ba
P^{QM}_{+\pm}(l)=\demi+\frac{1}{3}\cos(\alpha+\beta)\nonumber\\
P^{QM}_{-\pm}(l)=\demi-\frac{1}{3}\cos(\alpha+\beta)
\label{eq:SPQMl2}
\ea

Eq. (\ref{eq:SPQML2}) and (\ref{eq:SPQMl2}) together show that an
observer watching only the detectors D$_{1}$ cannot become aware in
the present of actions performed in the future of his light cone.
However, according to QM the coincidences measurement should
demonstrate such influences acting really backward in time. We are
in a quite similar situation as for the faster than light
influences involved in the Bell-experiments: in this case the
coincidences measurement demonstrates real faster-than-light
influences, even though these influences cannot be used for
superluminal telegraphing.\\

\section{Comments and Answers}

{\large\bf\sc Comment 1.}\,\ {\bf\em Paper \cite{as98.1} suffers
from what I believe to be a severe problem. The author examines the
probability of detection at a detector for one photon of a
correlated pair, where the two photons traverse different
interferometers. He then restrict attention to only those pairs
which have a given path difference for the two photons.
Unfortunately, there is absolutely no way of knowing what the path
difference is which the two photons have traversed and still have
the required interference. Ie, if the wave packets are sufficiently
short that the timing of the arrival at the detectors can be used
to determine that path difference, then the coherence of the
photons in each of the paths is too short to allow the photons
traversing different paths in the one detector to interfere. $(L,
LL)$ and $(l, lL)$ will then not interfere with each other, and the
expression (\ref{eq:JPQML}) above for the quantum probability will
not be correct. The correct one will be:

\be
|A_{\sigma\omega}(L,LL)|^2 + |A_{\sigma\omega}(l,Ll)+A_{\sigma\omega}(l,lL)|^2
\ee

which is a very different expression. In general, quantum mechanics
will always predict that the probabilities at detector 2 will be
totally insensitive to the measurements, or the paths followed for
particle 1, if those are unknown. There will be correlations
between outcomes of the measurement made on particle 1, but the
outcomes of measurements made on particle one will be independent
of what kinds of measurements are made on particle 2. In this case
the author has assumed that one can both determine the timing of
the two particles with sufficient accuracy to differentiate between
various possibilities, and at the same time have interference over
time scales which are longer than that time scale.}\\

This comment seems to overlook that the experiment involves two
different coherence lengths, that of the pump laser light (of about
30 m), and that of the photon pair light (of about 10 $\mu$m).\\

Time-resolved detection \cite{jb92,tbg97,ptjr94} of the photon
pairs cannot distinguish between the paths of subensemble $L$:
$(L,LL)$, $(l,lL)$, $(l,Ll)$, because all of them yield the same
time difference in the detected photon pair signals. Neither can
measurement of the time of emission of the pump laser light
distinguish between these paths when, as assumed, the path
difference $L-l$ (of about 0.3 m) between the long and the short
path of each interferometer is much smaller than the pump beam
coherence length. Therefore according to quantum mechanics the
paths of subensemble $L$ will interfere with each other, and for
the same reasons the paths of subensemble $l$: $(l,ll)$, $(L,lL)$,
$(L,Ll)$, will also interfere with each other too.\\

On the contrary, time-resolved detection allows us to discriminate
between the cases where a pair follows a path of subensemble $L$,
and the cases where the pair follows a path of subensemble $l$.
Therefore, if QM holds, a time delay spectrum of coincidence counts
\cite{jb92,ptjr94} for each of the four possible outcomes
D$_{1}(\sigma)$, D$_{2}(\omega)$, will exhibit four peaks: an
interference peak corresponding to subensemble $L$ we suppose set
at time difference 0, a second interference peak at time difference
$(l-L)/c$ corresponding to subensemble $l$, and two other peaks at
time differences $(L-l)/c$ and $(2l-2L)/c$ corresponding to path
$(l,LL)$, respectively $(L,ll)$. Using a time difference window one
can select only the events corresponding to subensemble $L$, or
only those to subensemble $l$.\\

In conclusion, there is no problem of principle to perform the
proposed experiment, and as regards the quantum mechanical
predictions, expression (\ref{eq:JPQML})  gives the correct quantum
mechanical probability.\\

{\large\bf\sc Comment 2.}\,\ {\bf\em Consider the case in which the
detections at the right-hand side of the setup, in detectors
D$_{2}$, lie time-like separated after the detections at the
left-hand side, in detectors in D$_{1}$. By selecting the paths
$(L,LL)$, $(l,Ll)$, $(l,lL)$ through a time interval between the
detection in one of the detectors D$_{1}$ and one of the detectors
D$_{2}$ you are using information from side 2 to select events on
side 1, and therefore determining the outcomes afterwards in
time.}\\

I don't think that this explanation reproduces the quantum
mechanical view. According to quantum mechanics one should
considered that each particle travels all the three
indistinguishable paths $(L,LL)$, $(l,Ll)$, $(l,lL)$ at once. This
means that:\\

\begin{enumerate}
\item{The joint probabilities $P^{QM}_{\sigma\omega}$ predicted by
the superposition principle result from contributions from these
three paths alone, and}
\item{These three paths have to be considered
absolutely equivalent to each other, i.e., each of them contributes
the same amount to the quantity $P^{QM}_{\sigma\omega}$.}
\end{enumerate}

\smallskip

From these conditions it follows that even if the measurement
selects only those counts in the detectors D$_{i}(\sigma)$ yielding
path difference $L$ through coincidence with the counts in the
detectors D$_{j}(\omega)$, the measured distribution of the
outcomes is the same as if it had been possible to perform the
experiment nonselectively, with only the three paths belonging to
the subensemble $L$ (what admittedly is not feasible).\\

But in the latter hypothetical case the causality principle
requires single detection probabilities for the photon detected
first depending only on the parameters it meets on its travel from
S to the detector. As said, the quantum mechanical predictions (Eq.
(\ref{eq:SPQML1}), (\ref{eq:SPQML2})) clearly violate this
requirement for Time ordering 2. Indeed the violation is plain
since an observer watching only detectors D$_{1}$ would measure (in
the hypothetical case) count rates (\ref{eq:SPQML2}) depending on
choices of parameter $\beta$ lying in his light-cone future, i.e.,
QM implies influences acting backward in time between time-like
separated regions.\\

Notice that from a quantum mechanical point of view the selection
of paths through detection time interval can very well be
interpreted as a reversed preparation -a {\em retroparation}-
\cite{co98}, and the fact that we are not capable to perform the
experiment using only the three paths referred to, is no argument
against the existence of influences backwards in time, but only
against the claim to use them for practical purposes, e.g. to build
a Time Machine. The similarity with superluminal nonlocality is
impressive, excepted obviously that Bell experiments have already
been done and upheld QM, whereas the experiments we are discussing
have not yet been done, and could in principle reject it.\\

{\large\bf\sc Comment 3.}\,\ {\bf\em A very causal and superluminal
interpretation holds for time ordering 2 (detections in D$_{2}$
time-like separated after detections in D$_{1}$): Detection of
photon 1 determines the possible paths photon 2 can choose. If
photon 1 is detected in $+$, then photon 2 is obliged to take a
path such that the time interval between detections is $L$;
conversely, if photon 1 is detected in $-$, then photon 2 is
obliged to take a path such that the time interval between
detections is $l$. I don't see any violation of causality (striking
or not) in this fact.}\\

Indeed the dependence on $\cos(\alpha+\beta)$ exhibited by the
single quantum mechanical probabilities $P^{QM}_{+\pm}$ at side 1
(left-hand side) of the setup (see Eq. (\ref{eq:SPQML2})) clearly
arise because in the interference at BS$_{22}$ the paths $(L,LL)$
and $(l,lL)$, which carry out half of the terms of the second order
interference at BS$_{21}$, do not interfere with the paths
$(L,Ll)$, $(l,ll)$ which carry out the other half. This description
assuming some kind of intermediate nonlocal correlations between
the outcomes at side 1 and the output ports of BS$_{21}$ is
undoubtedly a superluminal causal, but not a quantum mechanical
one. According to quantum mechanics it is detection what brings the
wavefunction to collapse, and it is not ortodox to assume a kind of
reduction before, i.e. you could if you wish assume the detection
value on side 1 to determine the detection value on side 2, but it
does not make sense (within the quantum mechanical view) to speak
about "the path by which the photon leaves BS$_{21}$" if it gets
detected only after BS$_{22}$.\\

Moreover it is clear that such an explanation with interferences in
between cannot account for the functioning of the quantum
mechanical superposition principle in the proposed experiment
because of the terms coming from path $(l,Ll)$. This path
interferes at BS$_{22}$ with the other two ones, but not at
BS$_{21}$. The presence of this path produces terms in
$\cos(\alpha+\gamma)$ which imply a direct nonlocal correlation
between detections at side 1, in the detectors monitoring
BS$_{11}$, and detections at side 2, in the detectors monitoring
BS$_{22}$. Accordingly, if detectors D$_{2}$ lie time-like
separated after detectors in D$_{1}$, a dependence on
$\cos(\alpha+\beta)$ of the single probabilities $P^{QM}_{+\pm}$ in
detectors D$_{1}$ could only be interpreted as a retrocausal
link.\\

{\large\bf\sc Comment 4.}\,\ {\bf\em Assumed the superposition
principle violates the causality principle in the proposed
experiment, it seems clear that a nonlocal causal description must
prescribe another rule to calculate the joint probabilities. In
\cite{as98.1} it is proposed to assume there is no superluminal
influence between side 1 and side 2. Why not to assume such
influences at least between BS$_{11}$ and BS$_{21}$?}\\

In a certain sense this has been done in \cite{as98.2}, for in this
paper a model was proposed assuming superluminal influences as well
between BS$_{11}$ and BS$_{21}$ for half of the particles traveling
$(L,LL)$ and half of the particles traveling $(l,lL)$, as between
BS$_{11}$ and BS$_{22}$ for half of the particles traveling
$(L,LL)$ and half of the particles traveling $(l,Ll)$. Meanwhile I
think the most reasonable assumption is to exclude superluminal
influences between BS$_{11}$ and BS$_{22}$ for all particles, and
accept them between BS$_{11}$ and BS$_{21}$ for the particles
traveling the paths $(L,LL)$ and $(l,lL)$.\\

Indeed nothing seems to speak against defining indistinguishability
in a stronger way than quantum mechanics does, and state that a
number of paths can produce interferences of a certain order at
BS$_{2l}$ only if all of them do interfere in any preceding
interference of the same order at BS$_{2k}$, $k<l$. According to
this assumption paths $(L,LL)$, $(l,Ll)$ and $(l,lL)$ cannot be
considered to interfere at BS$_{22}$ because two of them $(L,LL)$
and $(l,lL)$ do interfere nonlocally at BS$_{21}$, but the third
one $(l,Ll)$ does not. We call this principle {\em causal
indistinguishability} because the motivation to introduce it is to
unify superluminal influences and causality principle within the
frame of Relativistic Nonlocality or Multisimultaneity
\cite{asvs97.1, as97.2}. In the next section we derive
corresponding predictions.\\

We would like to point out that till the experiment proposed in
Fig. 1 there has been no necessity to define indistinguishability
in this more stronger way, for quantum mechanical
indistinguishability reduces to {\em causal indistinguishability}
in all setups considered before. This is also the case for the
great variety of possible experiments with series of {\em before}
and {\em non-before} impacts described in \cite{as97.1}.\\

{\large\bf\sc Comment 5.}\,\ {\bf\em Bell experiments with
time-like separated impacts at the splitters have already been done
\cite{ptjr94} demonstrating the same correlations as for space-like
separated ones. Cannot be said that such experiments, as well as
Wheeler's ``delayed-choice'' ones, already demonstrate influences
backward in time? }\\

Actually not. Regarding the experiment described in \cite{ptjr94}
Quantum Mechanics predicts single counts equally distributed for
the photon impacting before. But this is also the prediction of the
causal view according to which the photon impacting first chooses
its outcome without being influenced by the parameters the photon
impacting later will meet at the other arm of the setup. Therefore,
concerning the experiment of \cite{ptjr94} it is possible to
account for the quantum mechanical predictions and the observed
results by means of causal links acting forward in time.\\

Regarding Wheeler's ``delayed-choice'' experiments, they can be
easily explained assuming unobservable subluminal causality acting
forward in time, as for instance Bohm's causal model does, and the
principles proposed in \cite{as97.1} also suggest.\\

\section{Completing Multisimultaneity through
the 'Causal Indistinguishability Condition'}

We now extend the basic principles and theorems of RNL or
Multisimultaneity presented in previous work
\cite{asvs97.1,as97.2,as97.1,as97.3} to the impact series
experiments proposed above, and show the necessity of defining
indistinguishability in a stronger way. To this aim, we discuss
impact series with moving beam-splitters involving
Multisimultaneity (i.e. several simultaneity frames) and, as
particular cases, the two possible time orderings in the experiment
of Fig. 1 with beam-splitters at rest (i.e., involving only one
simultaneity frame).\\

Remember that in Multisimultaneity the outcome values for
detections after beam-splitter BS$_{ik}$ are considered determined
at the time of arrival at this beam-splitter, and not at the time
of arrival at the detectors monitoring the output ports of
BS$_{ik}$.\\

At time $T_{ik}$ at which particle $i\,(i\in\{1,2\})$ arrives at
beam-splitter BS$_{ik}$ we consider in the inertial frame of this
beam-splitter which beam-splitters BS$_{jl}$ particle $j\,
(j\in\{1,2\}, j\neq i)$ did already reach, i.e. we consider whether
the relation $(T_{ik}<T_{j1})_{ik}$ holds, or there is a BS$_{jl}$
such that the relation $(T_{jl}\leq T_{ik}<T_{jl+1})_{ik}$ holds,
the subscript $ik$ after the parenthesis meaning that all times
referred to are measured in the inertial frame of BS$_{ik}$.\\

Suppose first of all the case in which $(T_{11}<T_{21})_{11}$, and
$(T_{2k}<T_{11})_{2k}$, for each $k$. This means the impacts of
photon 1 on BS$_{11}$ in the inertial frame of this beam-splitter
to occur before the impacts of photon 2 on BS$_{21}$, and the
impacts of photon 2 on each BS$_{2k}$ in the inertial frame of
BS$_{2k}$ to occur before the impacts of photon 1 on  BS$_{11}$.
This obviously requires fast moving beam-splitters. All impacts are
then called {\em before} ones, and labeled $b_{ik}$.\\

In this case the relativistic nonlocal causal view of
Multisimultaneity states that the outcome choices at each
beam-splitter take account of local information only (excepted
obviously the information that the actual impact is a {\em before}
one), i.e., the distribution of the outcome values of photon $i$ in
beam-splitter BS$_{ik}$ are not influenced by the parameters the
other photon $j$ meets in side $j$ of the setup. \\

Since we consider photon pairs that travel the paths $(L,LL)$,
$(l,Ll)$, $(l,lL)$, from the point of view of BS$_{22}$ once photon
2 did impact, no ulterior detection makes it possible to
distinguish between the path segments $(lL)$ and $(Ll)$. On the
contrary one cannot yet exclude to know whether photon 2 traveled
path segment $(LL)$ by detecting particle 1 before it impacts on
BS$_{11}$, since this impact did not yet happen in the inertial
frame of BS$_{22}$. Therefore, from the point of view of BS$_{22}$
path segments $(lL)$ and $(Ll)$ lead to first order interferences,
and path segment $(LL)$ does not interfere at all. As regards
things considered from the point of view of BS$_{11}$'s inertial
frame one cannot exclude that photon 2 is getting detected before
it impacts on BS$_{21}$ and reveals whether photon 1 did travel the
long path $(L)$ or the short one $(l)$. Therefore no interference
takes place at BS$_{11}$. Thus one is led to the relation:

\ba
P(b_{11},b_{21},b_{22})_{\sigma\omega}
=|A_{\sigma}(L)|^2|A_{\omega}(LL)|^2
+|A_{\sigma}(l)|^2|A_{\omega}(Ll)+ A_{\omega}(lL)|^2,
\label{eq:Pbbb}
\ea

where $P(b_{11},b_{21},b_{22})_{\sigma\omega}$ denotes the joint
probability of getting photon 1 detected in  D$_{1}(\sigma)$ and
photon 2 in  D$_{2}(\omega)$ for the paths of subensemble $L$ when
all impacts at both sides are {\em before} ones, and
$A_{\sigma}(path)$, $A_{\omega}(path)$ are the first order
amplitudes for the indicate path segments.\\

Substituting into (\ref{eq:Pbbb}) according to (\ref{eq:ALl}),
(\ref{eq:AlL}), and (\ref{eq:ALL}) in the Appendix one gets the
following joint probabilities:

\be
P(b_{11},b_{21},b_{22})_{\sigma\omega}
=\frac{1}{4}+\omega\frac{1}{6}\cos(\beta-\gamma)\\
\label{eq:bbbjp}
\ee

\smallskip

Eq. (\ref{eq:bbbjp}) yields the following single probabilities for
side 2:

\be
P(b_{22})_{\omega}
=\frac{1}{2}+\omega\frac{1}{3}\cos(\beta-\gamma)\\
\label{eq:b22}
\ee

and for side 1:

\be
P(b_{11})_{\sigma}=\frac{1}{2}
\label{eq:b11}
\ee

i.e. as regard the outcome distribution for the single detectors
D$_{2}(\omega)$ when the impacts on BS$_{22}$ are {\em before} ones
Multisimultaneity agrees with the time ordering insensitive
predictions of Quantum Mechanics in (\ref{eq:SPQML1}). On the
contrary, as regard the outcome distribution for the single
detectors D$_{1}(\sigma)$  when the impacts on BS$_{11}$ are {\em
before} ones Multisimultaneity clearly conflicts with what Quantum
Mechanics predicts in (\ref{eq:SPQML2}), i.e.:

\be
P(b_{11})_{\sigma}\neq P^{QM}_{\sigma\pm}
\label{eq:b11qm1}
\ee

\bigskip

Consider now the situation in which $(T_{11}<T_{21})_{11}$,  and
$(T_{2k}\geq T_{11})_{2k}$, for each $k$. This situation is given
in the arrangement of Fig.1 with beam-splitters at rest and set in
place according to time ordering 2. In such experiments, can the
impacts on BS$_{21}$ and BS$_{22}$ be considered together to be
{\em non-before} ones?\\

They cannot, because detections after BS$_{11}$ and BS$_{21}$, and
detections after BS$_{11}$ and BS$_{22}$ do not share the same
subensemble of indistinguishable paths, or, in other words, from
the three paths $(L,LL)$, $(l,lL)$, $(l,Ll)$ interfering with each
other at BS$_{11}$ and BS$_{22}$, only two $(L,LL)$, $(l,lL)$
interfere with each other at BS$_{11}$ and BS$_{21}$.\\

In fact the assumption that these impacts are $a_{21}$, $a_{22}$,
i.e. {\em non-before} ones, bears oddities, since then according to
the principles of Multisimultaneity the sum-of-amplitudes rule
would apply, and therefore on the one hand it holds that:

\ba
P(b_{11},a_{21},a_{22})_{\sigma\omega}=P^{QM}_{\sigma\omega},
\label{eq:Pb11a21a22Pqm}
\ea

and on the other hand it holds that:

\ba
P(b_{11},a_{21},a_{22})_{\sigma\omega}
=P(b_{11})_{\sigma} P\Big((a_{21}, a_{22})_{\omega}|(b_{11})_{\sigma}\Big).
\label{eq:Pb11a21a22}
\ea

From Eq. (\ref{eq:Pb11a21a22Pqm}) and (\ref{eq:Pb11a21a22}) it
follows that:

\ba
P^{QM}_{\sigma\pm}= P(b_{11}, a_{21},a_{22})_{\sigma\pm}\nonumber\\
=P(b_{11})_{\sigma}\Big[P\Big((a_{21}, a_{22})_{+}(L)|(b_{22})_{\sigma}\Big)+
P\Big((a_{21},a_{22})_{-}|(b_{22})_{\sigma}\Big)\Big]\nonumber\\
=P(b_{11})_{\sigma}
\label{eq:Pqm=Pb11}
\ea

which contradicts Eq. (\ref{eq:b11qm1}).\\

This contradiction just means that in this experiment the
sum-of-amplitudes rule violates the causal view of
Multisimultaneity and, therefore the probabilities have to be
calculate otherwise.\\

If the impact on BS$_{21}$ and the subsequent impact on BS$_{22}$
cannot be both together {\em non-before} ones, then a quite
reasonable assumption can be to treat the first of them as {\em
non-before} impact for the paths $(L,LL)$, $(l,lL)$ interfering
with each other at BS$_{11}$ and BS$_{21}$, and the impact on
BS$_{22}$ as {\em before} one yielding only first degree
interferences for the paths $(l,lL)$ and $(l,Ll)$. More in general
we introduce the following interference condition:\\

{\em Causal Indistinguishability Condition}: A number of paths can
produce interferences of a certain order at BS$_{il}$ only if all
of them do interfere in any preceding interference of the same
order at BS$_{ik}$, $k<l$.\\

Taking account of this condition within Multisimultaneity one is
lead to the following rules:\\

\begin{enumerate}
\item{Path $(L,LL)$ and path $(l,lL)$ produce nonlocal 2-photon interferences
at BS$_{11}$ and BS$_{21}$.}
\item{Path $(l,Ll)$ and path $(l,lL)$ produce single photon interferences at BS$_{22}$.}
\item{Path $(L,LL)$ and path $(l,Ll)$ do not produce nonlocal 2-photon interferences
at BS$_{11}$ and BS$_{22}$.}
\end{enumerate}

\smallskip

Moreover as stated in \cite{as97.1} we assume all paths involved in
an interference to contribute the same way to the outcome
distribution, i.e. each path yields the same probability for a
determined outcome. According to these rules, the single paths give
the following contributions to the joint probabilities:

\ba
P(b_{11}, a_{21},b_{22})_{\sigma\omega}(L,LL)&=&\frac{1}{4}
\Big[1-\sigma\omega\cos(\alpha+\beta)\Big]\nonumber\\
P(b_{11}, a_{21},b_{22})_{\sigma\omega}(l,lL)&=&\frac{1}{4}
\Big[1-\sigma\omega\cos(\alpha+\beta)\Big]
\Big[1+\sigma\omega\cos(\gamma-\beta)\Big]\nonumber\\
P(b_{11}, a_{21},b_{22})_{\sigma\omega}(l,Ll)&=&\frac{1}{4}
\Big[1+\sigma\omega\cos(\gamma-\beta)\Big]
\label{eq:mcjpsp}
\ea

\smallskip

Equations (\ref{eq:mcjpsp}) yield the total joint probabilities:

\ba
P(b_{11},a_{21},b_{22})_{+ +}=\frac{1}{12}
\Big[3-2\cos(\alpha+\beta)+2\cos(\gamma-\beta)
-\cos(\alpha+\beta)\cos(\gamma-\beta)\Big]\nonumber\\
P(b_{11}, a_{21},b_{22})_{+ -}=\frac{1}{12}
\Big[3-2\cos(\alpha+\beta)-2\cos(\gamma-\beta)
+\cos(\alpha+\beta)\cos(\gamma-\beta)\Big]\nonumber\\
P(b_{11},a_{21},b_{22})_{- +}=\frac{1}{12}
\Big[3+2\cos(\alpha+\beta)+2\cos(\gamma-\beta)
+\cos(\alpha+\beta)\cos(\gamma-\beta)\Big]\nonumber\\
P(b_{11},a_{21},b_{22})_{- -}=\frac{1}{12}
\Big[3+2\cos(\alpha+\beta)-2\cos(\gamma-\beta)
-\cos(\alpha+\beta)\cos(\gamma-\beta)\Big]
\label{eq:mcjp}
\ea

\smallskip

From (\ref{eq:mcjp}) one gets the multisimultaneous causal
correlation coefficient:

\ba
E=\sum_{\sigma,\omega}(-\sigma\omega)
P(b_{11},a_{21},b_{22})_{\sigma\omega}&=&\frac{1}{3}\cos(\alpha+\beta)\cos(\gamma-\beta),
\label{eq:mccc}
\ea

the single probabilities for the left-hand side:

\ba
P(b_{11})_{\sigma}&=&\demi-\sigma\frac{1}{3}\cos(\alpha+\beta),
\label{eq:mcsp1}
\ea

and the single probabilities for the right-hand side:

\ba
P(b_{22})_{\omega}&=&\demi+\omega\frac{1}{3}\cos(\beta-\gamma).
\label{eq:mcsp2}
\ea

\smallskip

As regards other possible time orderings with beam-splitters at
rest, it is easy to see that for all of them the {\em Causal
Indistinguishability Condition} forbids second order interferences
at BS$_{22}$, and therefore the preceding Eq. (\ref{eq:mccc}),
(\ref{eq:mcsp1}), and (\ref{eq:mcsp2}) hold. Hence, contrarily to
the experiment proposed in \cite{as97.1}, the experiment of Fig. 1
with all beam-splitters resting in the laboratory frame does not
make possible to arrange a {\em non-before} impact at each side of
the setup and, therefore, is insensitive for different time
orderings.\\

As regards the time ordering in which the impacts on  BS$_{11}$ and
BS$_{21}$ become both {\em before} ones by means of fast moving
beam-splitters, no second order interference takes place at these
beam-splitters. Accordingly the {\em Causal Indistinguishability
Condition} does not forbid the impact on BS$_{22}$ to be a {\em
non-before} one. This $(b_{11}, b_{21}, a_{22})$ case is discussed
in a separated article, and shown that rules different from
conventional superposition and sum-of-probabilities may apply, as
it is the case for the corresponding $(b_{11}, b_{21}, a_{22})$
experiment discussed in \cite{as97.1}.\\

In summary, Eq. (\ref{eq:SPQML1}), (\ref{eq:SPQML2}), and
(\ref{eq:mcsp1}), (\ref{eq:mcsp2}), show that regarding single
detections at both sides the multisimultaneous model with Causal
Indistinguishability Condition yields the same single detection
values than Quantum Mechanics. On the contrary (\ref{eq:qmcc}) and
(\ref{eq:mccc}) show the two theories to conflict regarding the
correlation coefficient for joint probabilities.\\

Notice that RNL or Multisimultaneity, though causal, is a specific
superluminal nonlocal theory. That it conflicts with QM suggests
that the issues of superluminal nonlocality and of retrocausation
are not really entangled, and should be conceptually distinguished:
Nothing speaks in principle against the possibility that Nature
uses faster-than-light influences but avoids backward-in-time
ones.\\

\section{Real experiment}

A real experiment can be carried out arranging the setup used in
\cite{ptjr94} in order that the photon traveling the long fiber of
4.3 km impacts on a second beam-splitter before it is getting
detected. For phase values fulfilling the relations:

\ba
\alpha+\gamma=0,\nonumber\\
\alpha+\beta=\frac{\pi}{2},\nonumber\\
\beta-\gamma=\frac{\pi}{2},
\label{eq:se}
\ea

the equations (\ref{eq:qmcc}) and (\ref{eq:mccc}) yield the
predictions:

\ba
E^{QM}=\frac{2}{3}\nonumber\\
E=0
\label{eq:re}
\ea

\smallskip

Hence, for settings according to (\ref{eq:se}), e.g.,
$\alpha=45^{\circ}$, $\beta=45^{\circ}$, $\gamma=-45^{\circ}$, the
experiment represented in Fig. 1 allow us again to decide between
quantum mechanics and the multisimultaneous causal model proposed
above, through determining the corresponding experimental
quantities from the four measured coincidence counts
$R_{\sigma\omega}$ in the detectors.\\

\section{Conclusion}

The quantum mechanical predictions for the experiment represented
in Fig.1 with detections at side 2 lying time-like separated after
detections at side 1, cannot be reproduced through links forwards
in time, neither through such between the detectors, nor through
intermediate links between the beam-splitters BS$_{11}$ and
BS$_{21}$. Therefore upholding of quantum mechanics by experiment
would mean that we would have to learn living with influences
backward in time as we have learnt living with influences faster
than light.\\

The alternative case of experiment contradicting Quantum Mechanics
does not mean to give up influences faster than light, but only
influences backward in time arising from the way QM manages
indistinguishability and superposition. Superluminal nonlocality
and retrocausation are not necessarily entangled: in the frame of
Multisimultaneity, both superluminal causal links and the
impossibility of influences acting backwards in time have the
status of principles.\\

We would like to finish by stressing that the "overwhelming
experimental success" of quantum mechanics cannot be advanced as a
valid reason against doing the proposed impact series experiments.
Effectively in all experiments already done both concepts,
conventional quantum mechanical indistinguishability and the new
{\em Causal Indistinguishability Condition} proposed above, are
indistinguishable from each other so that the "overwhelming
experimental success" is so far shared by both. Moreover causal
indistinguishability looks to be a more sharply defined concept
than the quantum mechanical one. Therefore, in order to distinguish
the way nature does really work, it is necessary to perform
experiments in which both indistinguishabilities become
distinguishable. Whatever the answer may be, impact series
experiments seem capable of bearing a promising controversy between
QM and Causality, similar to the controversy between QM and Local
Realism.\\

\section*{Acknowledgements}

I would like to thank Valerio Scarani (EPFL, Lausanne) and Wolfgang
Tittel (University of Geneva) for numerous suggestions, and Olivier
Costa de Beauregard (L. de Broglie Foundation, Paris) for
stimulating discussions on retrocausation. It is a pleasure to
acknowledge also discussions regarding experimental realizations
with Nicolas Gisin and coworkers (University of Geneva), and
support by the L\'eman and Odier Foundations.

\section*{Appendix}

In the following are listed the probability amplitudes of the path
pairs and the single paths we are interested in.

\subsection*{A.1\,Probability Amplitudes of the path pairs with
length difference $L$ in Table (\ref{eq:paths})}

We denote $A_{\sigma\omega}(\mbox{path})$ the probability amplitude
associated to detection of photon 1 in $D_{1}(\sigma)$ and of
photon 2 in $D_{2}(\omega)$, for the specified path. The
probability amplitudes for the path pairs of subensemble $L$ in
(\ref{eq:paths}) normalized to only these three path pairs are:

\ba
(l,Ll)&:&\left\{\begin{array}{lllll}
A_{++}(l,Ll)&=&-\,A_{--}(l,Ll)&=&-\frac{1}{\sqrt3}\,\frac{1}{2}\,e^{i\beta}\\
A_{+-}(l,Ll)&=&A_{-+}(l,Ll)&=&-i\,\frac{1}{\sqrt3}\,\frac{1}{2}\,e^{i\beta}
\label{Al,Ll}
\end{array}
\right.
\\
(l,lL)&:&\left\{\begin{array}{lllll}
A_{++}(l,lL)&=&A_{--}(l,lL)&=&-\frac{1}{\sqrt3}\,\frac{1}{2}\,e^{i\gamma}\\
A_{+-}(l,lL)&=&-\,A_{-+}(l,lL)&=&i\,\frac{1}{\sqrt3}\,\frac{1}{2}\,e^{i\gamma}
\label{Al,lL}
\end{array}
\right.
\\
(L,LL)&:&\left\{\begin{array}{lllll}
A_{++}(L,LL)&=&-\,A_{--}(L,LL)&=&\frac{1}{\sqrt3}\,\frac{1}{2}\,e^{i(\alpha+\beta+\gamma)}\\
A_{+-}(L,LL)&=&A_{-+}(L,LL)&=&-i\,\frac{1}{\sqrt3}\,\frac{1}{2}\,e^{i(\alpha+\beta+\gamma)}
\end{array}
\right.
\label{AL,LL}
\ea

\subsection*{A.2 \,Probability Amplitudes of the path pairs with
length difference $l$ in Table (\ref{eq:paths})}

The probability amplitudes for the path pairs of subensemble $l$ in
(\ref{eq:paths}) normalized to only these three path pairs are:

\ba
(l,ll)&:&\left\{\begin{array}{lllll}
A_{++}(l,ll)&=&-\,A_{--}(l,ll)&=&\frac{1}{\sqrt3}\,\frac{1}{2}\\
A_{+-}(l,ll)&=&A_{-+}(l,ll)&=&\frac{1}{\sqrt3}\,\frac{1}{2}\,i
\label{Al,ll}
\end{array}
\right.
\\
(L,lL)&:&\left\{\begin{array}{lllll}
A_{++}(L,lL)&=&-\,A_{--}(L,lL)&=&\frac{1}{\sqrt3}\,\frac{1}{2}\,e^{i(\alpha+\gamma)}\\
A_{+-}(L,lL)&=&A_{-+}(L,lL)&=&-\,\frac{1}{\sqrt3}\,\frac{1}{2}\,i\,e^{i(\alpha+\gamma)}
\label{AL,lL}
\end{array}
\right.
\\
(L,Ll)&:&\left\{\begin{array}{lllll}
A_{++}(L,Ll)&=&A_{--}(L,Ll)&=&\frac{1}{\sqrt3}\,\frac{1}{2}\,e^{i(\alpha+\beta)}\\
A_{+-}(L,Ll)&=&-\,A_{-+}(L,Ll)&=&\frac{1}{\sqrt3}\,\frac{1}{2}\,i\,e^{i(\alpha+\beta)}
\label{AL,Ll}
\end{array}
\right.
\ea

\subsection*{A.3\,Probability Amplitudes of the single path segments $LL$, $Ll$, $lL$
traveled by photon 2}

We denote $A_{\sigma}(\mbox{path})$ the probability amplitude
associated to detection of photon 2 in D$_{2}(\sigma)$, for the
specified path segment. The probability amplitudes for the path
segments $LL$, $Ll$, $lL$ photon 2 travels in an experiment
selecting the path pairs with path difference $L$ in
(\ref{eq:paths}), normalized as if the experiment were performed
with only these three paths are:

\ba
(Ll)&:&\left\{\begin{array}{lll}
A_{+}(Ll)&=&-\frac{1}{\sqrt3}\frac{1}{\sqrt2}\,e^{i\beta}\\
A_{-}(Ll)&=&-i\,\frac{1}{\sqrt3}\frac{1}{\sqrt2}\,e^{i\beta}
\label{eq:ALl}
\end{array}
\right.
\\
(lL)&:&\left\{\begin{array}{lll}
A_{+}(lL)&=&-\frac{1}{\sqrt3}\frac{1}{\sqrt2}\,e^{i\gamma}\\
A_{-}(lL)&=&i\,\frac{1}{\sqrt3}\frac{1}{\sqrt2}\,e^{i\gamma}
\label{eq:AlL}
\end{array}
\right.
\\
(LL)&:&\left\{\begin{array}{lll}
A_{+}(LL)&=&-\frac{1}{\sqrt3}\frac{1}{\sqrt2}\,e^{i(\beta+\gamma)}\\
A_{-}(LL)&=&i\,\frac{1}{\sqrt3}\frac{1}{\sqrt2}\,e^{i(\beta+\gamma)}
\label{eq:ALL}
\end{array}
\right.
\ea

\pagebreak

\end{document}